\documentclass[twocolumn,showpacs,fleqn,nobibnotes]{revtex4}
\usepackage{amsmath}
\usepackage{graphicx}
\usepackage{float}
\usepackage{subfigure}
\usepackage{verbatim}
\usepackage{color}
\usepackage{comment}
\usepackage{hyperref}

\newcommand{\rr}{\mbox{\boldmath $r$}}

\begin{document}
	
	\title{Coherent and incoherent production of vector mesons in  ultraperipheral collisions of Xenon-ions within the QCD parton saturation approach}
	\author{F. Kopp, M.V.T. Machado}
	
	\affiliation{High Energy Physics Phenomenology Group, GFPAE  IF-UFRGS \\
		Caixa Postal 15051, CEP 91501-970, Porto Alegre, RS, Brazil}

\begin{abstract}
In this paper we analyse the exclusive vector meson photoproduction in the recent run using Xenon-ions at energy of 5.44 TeV performed by the Large Hadron Collider. We focus on the ultraperipheral collisions and provide theoretical predictions for coherent and incoherent cross sections within the color dipole approach and gluon saturation framework.  The rapidity distribution is investigated in both cases and comparison to other approaches available at literature is done. We show that the expected yields are enough to perform reliable cross section measurements for light mesons as $\rho^0$ and $\phi$.

\end{abstract}	

\maketitle
	
\section{Introduction}	

Investigating the exclusive meson photoproduction in ultraperipheral collisons (UPCs) \cite{upcs} is an essential tool to understand the underlying   dynamics of strong interactions. In the case of heavy meson production, it sheds light in the low-x physics and helps to constraint the nuclear gluon density specially at large meson rapidities, $y$.  As an example, in $J/\psi$  photonuclear production at the Large Hadron Collider (LHC) regime one obtains the value $x=\frac{m_{\psi}}{\sqrt{s_{AA}}}e^{-y}\simeq 3\times 10^{-5}$ at $y=3$ for PbPb collisions. In the UPCs, the nuclear target is probed by quasi-virtual photons and the typical momentum scale is the meson mass. For light mesons, the mass scale lies below 1 GeV and usual methods of weak coupling are not valid. In addition, even for the heavy meson production the mass value is typically associated to a semi-hard scale below 10 GeV. These features turn out the thoeretical approaches considering parton saturation very appealing. The reason is that the typical momentum scale at very low-$x$ would be the nuclear saturation scale, $Q_{sat,A}^2\simeq A^{1/3}Q_{sat,p}^2\sim 6$ GeV$^2$ (for Pb at very low-$x$), which is enhanced in case of large nuclei. A demonstration of the power of predictability of the parton saturation approach including the geometric scaling phenomenon has been presented in Ref. \cite{Ben:2017xny}, where the exclusive production of $\rho$ and $J/\psi$ is studied in photon-nuclei interactions. Accordingly, the systematics of strong nuclear amplification of gluon saturation from production of mesons in $eA$ collisions is presented in Ref. \cite{Mantysaari:2017slo}.  The photon-target interaction amplitude, when considering the light-cone dipole formalism \cite{nik}, can be written as a convolution between the photon-meson wave functions overlap and the elementary dipole-target cross section \cite{Nemchik:1996pp}. Within the color dipole approach one can introduce information on dynamics beyond the leading
logarithmic QCD approach and computing predictions for the radially excited states is a simple task \cite{Nemchik:1996pp}. From the experimental point of view, there is intense activity on measuring the rapidity and momentum transfer distributions for coherent and incoherent processes and the theoretical models in general consider distinct dynamics for low and high mass production. On the other hand, the parton saturation framework describes both  regimes in a unified way (for a recent study describing light and heavy meson photoproduction at the LHC, see for instance Ref. \cite{Goncalves:2017wgg}). 

In this work, we investigate the coherent and incoherent vector meson photoproduction in Xe+Xe collisions at the LHC for the energy of 5.44 TeV  per nucleon. This is motivated by the test run of collisions of Xenon ions  recently performed at LHC, which reached a statistics of several $\mu b^{-1}$. 
 Xenon (Xe) is around 40\% lighter than that of lead (Pb) and the QGP-like medium created in Xe+Xe collisions would be cooler and shorter lived when compared to PbPb collisions. The expectations are that differences observed in the measurements from the collisions of these two nuclei could provide valuable information on the underlying physics of nuclear environment. In order to predict the cross section for Xe+Xe collisions we will consider the QCD color dipole approach and phenomenological models including parton saturation phenomenon. Such an analysis has been previously considered for Pb+Pb collisions at the LHC with relative success. For instance, the theoretical uncertainty associated to the light meson production was investigated \cite{Santos:2014vwa}, where ALICE data for $\rho $ photoproduction have been described. Similar analysis was done \cite{Santos:2014zna} also for heavy mesons, including prediction for the incohent cross section for $J/\psi$ and $\Upsilon$. The detailed analysis related to the radial excitations of charmonia was investigated in Refs. \cite{Ducati:2017bzk,Ducati:2016jdg,Ducati:2013bya}, where the momentum transfer distribution has been also addressed.

Our analysis here is complementary to the studies done in Ref. \cite{Guzey:2018bay}, which  considers the photonuclear production of mesons $\rho$ and $\phi$  in the context of UPCs using Xenon ions and the corresponding role of structure factors of Xe isotopes in
Xenon-based detectors of Dark Matter (WIMP candidates in direct dark matter searches). The theoretical approach considered was the combination
 of Glauber-Gribov model to describe nuclear effects (shadowing) and a model for hadronic fluctuations for the photon-nucleon cross section, which is 
in agreement with the experimental results for $\rho$-photoproduction at RHIC (AuAu) and LHC (PbPb runs) energies. For the incoherent case, the authors analyse the size of target nucleon dissociation contribution in details. We consider it is timely and important to compare the the distinct approaches presented in literature and to estimate the size of theoretical uncertainties for the cross sections in the vector meson production.

The paper is organized as follows. In the next section we give the main theoretical information to obtain the rapidity distribution of coherent and incoherent production of vector mesons in Xe+Xe collisions at centre-of-mass energy of 5.44 TeV. Very recent experimental analysis for these collisions are ongoing, see for instance Refs. \cite{Acharya:2018jvc,Acharya:2018eaq,Acharya:2018ihu,CMS1,CMS2,CMS3,ATLAS1,ATLAS2}. We will demonstrate that the expected yields are enough to perform reliable cross section measurements for light mesons as $\rho^0$ and $\phi$. In the section \ref{discussions} we present the phenomenological calculations, discuss the main theoretical uncertainties and comparison with other approaches is performed.  Finally, we summarize the main conclusions.

\section{Theoretical Formalism}

Exclusive meson photoproduction in nucleus-nucleus collisions is factorized in terms of the equivalent flux of photons of the nucleus projectile and photon-target production cross section \cite{upcs}. In UPCs there is the absence of strong interactions between the projectile particle and the target and they are characterized by impact parameter $>$ 2 $R_{A}$. The interaction is ultra-relativistic and purely electromagnetic and in general one uses the Weizs\"{a}cker-Williams approximation \cite{upcs}. The photon energy spectrum, $dN_{\gamma}^A/d\omega$, which depends on the photon energy $\omega$, takes part in the expression  for the rapidity distribution of vector mesons ($V$) which is  written in the following way,
\begin{eqnarray}
\frac{d\sigma}{dy} (A A \rightarrow   A\otimes V \otimes Y) & = & \left[ \omega \frac{dN_{\gamma}^A}{d\omega }\,\sigma(\gamma A \rightarrow V\,Y ) \right. \nonumber \\
& + & \left. \left(y\rightarrow -y \right) \right],
\label{dsigdyA}
\end{eqnarray}
where $Y=A$ (coherent case) or $Y=A^*$ (incoherent case). The symbol $\otimes$ denotes the large rapidity gap between the produced meson and the final states nucleus.

The produced state with mass $m_V$ has rapidity $y\simeq \ln (2\omega/m_V)$ and the square of the $\gamma A$ centre-of-mass energy is given by $W_{\gamma A}^2\simeq 2\omega\sqrt{s}$. As one has a diffractive process, the exchanged object in the interaction carries vacuum quantum numbers (the Pomeron). Here, we consider the QCD realization of color singlet object based on the two-gluon  exchange and further gluon emmisions. That is, the photon-Pomeron interaction will be described within the light-cone dipole frame, where the probing
projectile fluctuates into a
quark-antiquark pair with transverse separation
$r$ (and momentum fraction $z$) long after the interaction, which then
scatters off the hadron. The cross section for exclusive photoproduction of the meson states  off a nucleon target is given by,
\begin{eqnarray}
\sigma (\gamma p\rightarrow V \,p) = \frac{\left|\sum_{h, \bar{h}} \int dz\, d^2r \,\Psi^\gamma_{h, \bar{h}}\sigma_{dip}(x,r)\, \Psi^{V*}_{h, \bar{h}}  \right|^2}{16\pi B_{V}} ,
\label{sigmatot}
\end{eqnarray}
where $\Psi^{\gamma}$ and $\Psi^{V}$ are the light-cone wavefunction  of the photon  and of the  vector meson, respectively (we will consider in this work the following states: $V = \rho,\,\phi,\,J/\psi,\,\psi(2S),\,\Upsilon(1S),\,\Upsilon (2S)$). The quark and antiquark helicities states are labelled by $h$ and  $\bar{h}$ , respectively. The dipole-proton cross section is denoted by  $\sigma_{dip}(x,r)$ and the  diffractive slope parameter by $B_V$.  In this context, we are implicitly assuming that the
proton shape is Gaussian and that the impact parameter dependence
factorizes out from the dipole-nucleon scattering amplitude. 

The exclusive photoproduction off nuclei for coherent and incoherent processes is computed in high energies (in the limit of large coherence length $l_c\gg R_A$) as follows \cite{Boris},
\begin{eqnarray}
\sigma (\gamma A \rightarrow V A) & = & K^2\int d^2b\, \left| \sum_{h, \bar{h}} \int dz\, d^2r \,\Psi^\gamma_{h, \bar{h}} \Psi^{V*}_{h, \bar{h}} \right.  \\
&\times & \left. \left[1-\exp\left(-\frac{1}{2}\sigma_{dip}(x,r) T_A(b)\right)  \right]\right|^2, \label{eq:coher} \nonumber \\
\sigma (\gamma A \rightarrow V A^* )  & = & K^2 \int d^2b\,\frac{T_A(b)}{16\pi\,B_V}\left| \sum_{h, \bar{h}} \int dz\, d^2r \,\Psi^\gamma_{h, \bar{h}} \Psi^{V*}_{h, \bar{h}} \right. \nonumber \\
 &\times & \left. \sigma_{dip}(x,r) \exp\left[-\frac{1}{2}\sigma_{dip}(x,r)T_A(b)  \right]\right|^2,
\label{eq:incoh}
\end{eqnarray} 
where $T_A(b)= \int dz\rho_A(b,z)$  is the nuclear thickness function. We performed the calculations with the two-parameter Fermi distribution of the nuclear density, $\rho (r) = \rho_0[1+\exp (r-R_A)/a)]^{-1}$, with $\rho_{0}=0.148$ fm$^{-3}$, $R_A=6.624$ fm and $a=0.54$ fm. Concerning the incoherent case, here we will not consider neutron emission. In distinction to coherent diffraction, the nucleus is allowed to break up, but except for the vector meson no new particles are produced in the reaction. We quote Ref. \cite{Luszczak:2017dwf} where the multiple scattering expansion of the incoherent diffractive cross section is derived as an expansion over quasielastic scatterings of the color dipole. Our expression in Eq. (\ref{eq:incoh}) corresponds to the first order term and should be the dominant contribution at small-$t$.

In the numerical evaluations, we will use boosted Gaussian wavefunction and several phenomenological saturation models, which encode the main properties of the QCD parton saturation approach.  The cross sections above should include both the skewedness and real part of amplitude corrections and we multiply the expressions by $K^2 = R_g^2 (1 + \beta^2)$, where $\beta = \tan (\pi \lambda_{ef}/2)$ is the ratio of real to imaginary parts of the scattering amplitude and $R_g $ incorporates the off-forward correction (see e.g. Refs. \cite{Santos:2014vwa,Santos:2014zna,Ducati:2016jdg} for details). The effective power on energy, $\lambda_{ef}$ is determined for each case. In order to take into account the threshold correction for the dipole cross section, we have multiplied them by a factor $(1-x)^n$ (with $n=5$ for light mesons and $n=7$ for the heavy ones).

Now, we set the parameters and phenomenological models to be considered in next section. For the slope parameter one considers an energy dependence based on Regge phenomenology \cite{Santos:2014vwa,Santos:2014zna,Ducati:2016jdg},
\begin{eqnarray}
B_{V}=b^{V}_{el}+2\alpha^{\prime}\text{log}\left(\frac{W^2_{\gamma A}}{W^2_0}\right).
\label{diffslope}
\end{eqnarray}
We call attention that $B_V$ is considered only for calculating the incoherent cross section. For the meson wavefuntion, one takes the Boosted-Gaussian model \cite{wfbg} because it can be applied in a systematic way for excited states.   It works well for the light mesons and also for the heavier mesons and it is  given by
\begin{eqnarray}
\psi_{\lambda,h\bar{h}}^{V}(z,r) & = &  \sqrt{\frac{N_{c}}{4\pi }}\frac{\sqrt{2}}{z(1-z)} \{\delta_{h,\bar{h}}\delta_{\lambda,2h}m_{q}  + i(2h)\delta_{h,-\bar{h}}\text{e}^{i\lambda\phi} \nonumber \\
& \times & \left[
(1-z)\delta_{\lambda,-2h}+z\delta_{\lambda,2h}\right]\partial_{r} \} \phi_{nS}(z,r)~,
\end{eqnarray}
with $\lambda$ being the meson helicity and where $\phi_{nS}(z,r)$ is given by \cite{Sanda2},
\begin{eqnarray}
\phi_{nS}(r,z) = \left[\sum^{n-1}_{k=0}\alpha_{nS,k}R^2_{nS}\hat{D}^{2k}(r,z)\right]G_{nS}(r,z),
\end{eqnarray}
with $\alpha_{nS,0}=1$. The operator $\hat{D}^{2}(r,z)$ is defined by
\begin{eqnarray}
\hat{D}^{2}(r,z) = \frac{m_f^2-(\frac{1}{r}\partial_r+\partial^2_r)}{4z(1-z)}-m_f^2,
\end{eqnarray}
 and it acts on the following generatrix function 
\begin{eqnarray}
 G_{nS}(r,z) & = & \mathcal{N}_{nS}\,z(1-z)\,\exp\left(-\frac{m^2_f\mathcal{R}^2_{nS}}{8z(1-z)} \right. \nonumber \\
& - & \left. \frac{2z(1-z)r^2}{\mathcal{R}^2_{nS}}+\frac{m^2_f\mathcal{R}^2_{nS}}{2}\right).
\end{eqnarray}

For the sake of completeness, in Tables \ref{tab:bslope} and \ref{tab:bGparams} are presented the parameters considered for the diffractive slope and parameters for the meson wavefunction, respectively.

\begin{table}[t]
	\centering
	\begin{tabular}{|c|c|c|c|}
		\hline \hline
		Meson   & $B_{0}$ [GeV$^{-2}$]   & W$_{0}$  [GeV]   & $\alpha^{\prime} $ [GeV$^{-2}$]   \\ \hline
		$\rho , \phi$  &      11.0            &       95.0    &          0.25       \\ \hline
		$J/\psi$ &      4.99           &       95.0    &          0.25       \\ \hline
		$\psi(2S)$ &      4.31           &       95.0    &          0.25       \\ \hline
		$\Upsilon(1S)$&      3.68           &       95.0    &          0.164       \\ \hline
		$\Upsilon(2S)$ &      3.61           &       95.0    &          0.164       \\ \hline

		\hline\hline
	\end{tabular}
	\caption{The parameters for the diffractive $B_V$ slope parameter, Eq. (\ref{diffslope}), for mesons considered in present work. }
	\label{tab:bslope}
\end{table}

We will use different models for the dipole scattering cross section: GBW \cite{PRD59-014017}, b-CGC \cite{PLB590-199} and IP-Sat \cite{PRD74-074016}. The GBW model is defined by an eikonal shape for the dipole cross section,
\begin{eqnarray}
\sigma^{GBW}_{q\bar q}(x,r)=\sigma_0\left(1-e^{-r^2Q^2_s(x)/4}\right)\label{gbw},
\end{eqnarray}
where $\sigma_0=2\pi R^2$ is a constant and $Q_s^2(x)=(x_0/x)^{\lambda}$ GeV$^2$ denotes the saturation scale. The parameters are fitted from DESY-HERA data and their values are $x_{0}=1.1 \cdot 10^{-4}$, $\lambda=0.287$, $\sigma_{0}$= 23.9 mb.  We also consider the b-CGC model \cite{PLB590-199}, based in the Color Glass Condensate framework, in which gluon saturation effects are incorporated via an approximate solution of the Balitsky-Kovchegov equation \cite{CGC}.  The expression for the b-CGC model is given by,
\begin{equation} \label{eq:bcgc}
\sigma_{q\bar{q}}^{BCGC}(x,r)=2\int d^2b \,\begin{cases}
\mathcal{N}_0\left(\frac{rQ_s}{2}\right)^{\gamma_{eff}(x,r)} & :\quad rQ_s\le 2\\
1-\mathrm{e}^{-A\ln^2(BrQ_s)} & :\quad rQ_s>2
\end{cases},
\end{equation}
where the parameter $Q_s$ now depends on the impact parameter:
\begin{equation} \label{eq:bcgc1}
Q_s\equiv Q_s(x,b)=\left(\frac{x_0}{x}\right)^{\frac{\lambda}{2}}\;\left[\exp\left(-\frac{b^2}{2B_{\rm CGC}}\right)\right]^{\frac{1}{2\gamma_s}},
\end{equation}
where $\gamma_{eff}(x,r)=2\left(\gamma_s+(1/\kappa\lambda \,\ln (1/x))\ln(2/rQ_s)\right)$ is the effective anomalous dimension and one has the constant $\kappa=9.9$.  The remaining parameters are ${\cal{N}}_{0}$=0.417, $x_{0}=5.95 \cdot 10^{-4}$, $\lambda= 0.159$, $\gamma_{s}=0.63$ and $B_{CGC}=5.5$ GeV.

Both GBW and b-CGC models do not include DGLAP evolution in their formulation. 
 In order to investigate the theoretical uncertainty involved in the different evolutions in saturation models, we also consider the IP-Sat model. The dipole cross section is now given by \cite{PRD74-074016,Amir},
\begin{eqnarray}
\sigma_{dip}^{IP-Sat}(x,\rr) & = & 2 \int d^2{\bf b}\, N(x,\rr, {\bf b}),  \\
 N(x,\rr, {\bf b}) & = & 1-\exp \left(-\frac{\pi^2r^2 \alpha_{s}(\mu^2) xg(x,\mu^2)T(b)}{2N_{c}}\right). \nonumber 
\end{eqnarray}
Here, $xg(x,\mu_0^2) = A_g\,x^{-\lambda_g}\,(1-x)^{5.6}$ with $\mu^2= \frac{ 4}{r^{2}} + \mu_0^2$. The profile function of the proton is considered as a gaussian, $T_G(b) = \frac{1}{2\pi B_G}\mathrm{e}^{-\frac{b^2}{2B_G}}$, where $B_G=4.0$ GeV$^{-1}$ is obtained from the fit of $\frac{d\sigma}{dt}$ for vector meson production \cite{PRD74-074016}. The remaining parameters are $\mu_{0} = 1.17$ GeV$^{2}$, $A_{g}=.55$, $\lambda_{g} $=0.020 and $\Lambda_{QCD}$=0.2 GeV.	

The quantity $xg(x,\mu_0^2)$ is the input at initial scale $ \mu_0^2=Q^{2}_{0}$  for DGLAP evolution without quarks, since the interaction between the nucleon and color dipole  occurs through the exchange of two gluons or one Pomeron. That is, the QCD evolution is given by,
\begin{equation}
\frac{\partial g(x,Q^2)}{\partial {\rm log}Q^2}~=~\frac{\alpha_s}{2\pi}~\left( \sum_i P_{gq} \otimes (q_i+\bar{q}_i)~+~P_{gg} \otimes g \right),
\end{equation}
 where QCDNUM\cite{qcdnum} was used for the DGLAP evolution.

\begin{table}[t]
	\centering
	\begin{tabular}{|c|c|c|c|c|c|c|}
\hline \hline
Meson   & $M_V$/GeV   & $m_f$/GeV  & $\mathcal{N}_T$   &  $\mathcal{R}^2$/GeV$^{-2}$ &
$\alpha_{s_{1}}$ & $e_{f}$ \\ \hline

$J/\psi$\cite{Armesto:2014sma}& 3.096       &   1.27         &     0.596              &  2.45        &             -        &              2/3       \\ \hline
$\psi$(2S)\cite{Armesto:2014sma}& 3.686     &      1.27      &       0.7            &   3.72       & -0.61                     &    2/3                 \\ \hline
$\phi$\cite{PRD74-074016}   &  1.019     &  0.14      & 0.919             &  11.2    &             -        & 1/3   \\ \hline
$\rho$\cite{PRD74-074016}   & 0.776      &  0.14      & 0.911             &  12.9    &             -         & 1/$\sqrt{2}$  \\ \hline
$\Upsilon$(1S)\cite{Sanda2} & 9.46 &  4.2      & 0.481             &  0.567    &             -          &   1/3   \\ \hline
$\Upsilon$(2S)\cite{Sanda2} & 10.023&  4.2      & 0.624             &  0.831    &   -0.555                     &   1/3   \\ \hline

\hline\hline
	\end{tabular}
	\caption{Parameters of the boosted Gaussian wavefunction, including the quark masses.}
	\label{tab:bGparams}
\end{table}

In next section we address the numerical calculation of coherent and incoherent cross sections in Xe+Xe collision at the energy of 5.44 TeV and investigate the theoretical uncertainty within the color dipole approach using the distinct implementations of the parton saturation phenomenon.

\section{Results and discussions}
\label{discussions}

	\begin{figure*}[t]
	\begin{tabular}{ccc}	
	\includegraphics[scale=1.0]{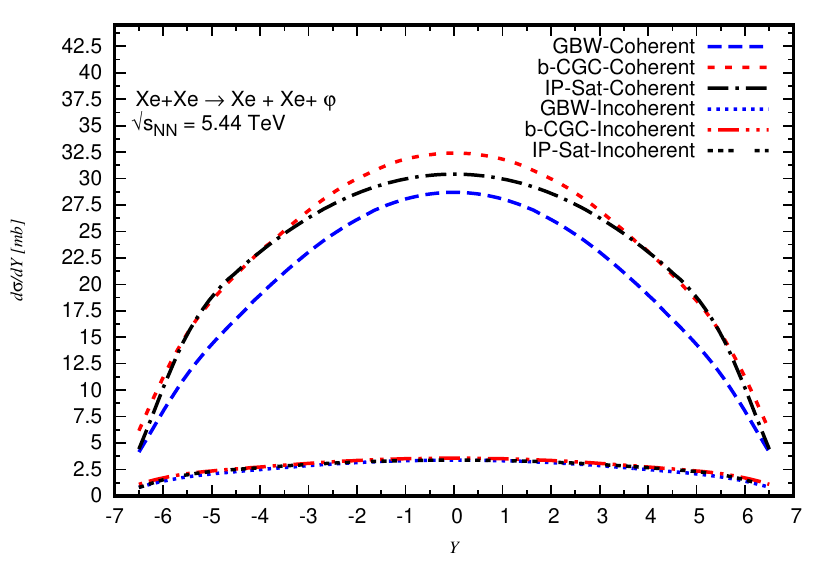} &
	\includegraphics[scale=1.0]{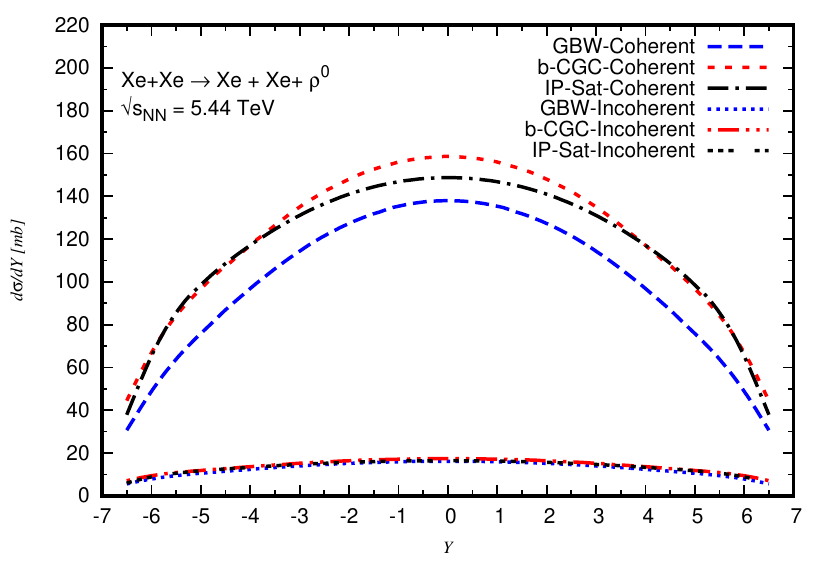} &   
		\end{tabular}
		\caption{Rapidity distributions for the  coherent (upper curves) and incoherent (lower curves) photonuclear production of $\phi$ and $\rho$ considering GBW, b-CGC and IP-Sat color dipole models (including parton saturation phenomenon).}
		\label{fig:1}
	\end{figure*}

Before calculating the cross sections we consider the present status of experimental feasibility for the processes we are considering here. The CMS Collaboration has measured Xe+Xe collisions with an integrated luminosity of 3.42 $\mu b^{-1}$ and the rapidity interval $|\eta | < 1$, whereas ATLAS Collaboration has measured 3 $\mu b^{-1}$ with $|\eta | < 2.5$. On the other hand, ALICE Collaboration has an limited integrated luminosity of 0.34 $\mu b^{-1}$ but presents an interesting analysis on $J/\psi$ production at forward rapidities, $2.5<y<4$. Given the typical efficiency for coherent vector meson production of a few percent (times the corresponding branching ratios) the expected yields should be enough to perform reliable $\rho$ or $\phi$ cross section measurements. However, the current statistics for Xe+Xe UPCs is too low in order to investigate quarkonia events. We will present the estimates for heavy mesons for the sake of completeness having in mind it can be useful for a long time future run using Xenon-ions.

We show in the following the theoretical predictions for the coherent (upper curves) and incoherent (lower curves)  processes for Xe+Xe UPCs at 5.44 TeV. The analysis is focused on the rapidity distribution and theoretical uncertainty from distinct dipole cross sections. In Figure $\ref{fig:1}$, it is presented the results for photonuclear production of $\rho$ (right panel) and $\phi$ (left panel) states, taking into account the different models presented in the last section. The theoretical uncertainty is  sizable being of order 6-7 \% for the both mesons in the coherent case. There would be an additional uncertainty related to the vector meson wave function (for details on the size of the related uncertainty, see e.g. Ref. \cite{Goncalves:2017wgg}).  At central rapidity one predicts explicitly $d\sigma_{coh} /dy\, (y=0)= 148.3\pm 10.3$ mb and $d\sigma_{coh} /dy\, (y=0)= 30.5\pm  1.8$ mb for $\rho$ and $\phi$, respectively. In addition, we get $d\sigma_{inc} /dy\, (y=0)= 8.4\pm 0.3\,\,(  1.73 \pm 0.05 )$ mb for $\rho \,(\phi)$ for the incoherent case, which has a smaller (around 3 \%) theoretical uncertainty that the coherent cross section. The relative contribution of the coherent $\rho$ compared to the $\phi$ states is $(\frac{d \sigma_{\rho}}{dy})/(\frac{d \sigma_{\phi}}{dy})=4.9 \pm 0.4 $ at $y=0$.   In order to study the feasibility of measurements, the typical efficiency for coherent $\rho$ in ALICE \cite{RhoALICE} is around $\bar{\epsilon}=(\mathrm{Acc}\times \varepsilon)_{\rho}\sim 7$ \% and the braching ratio $\mathrm{Br}(\rho \rightarrow \pi^⁺\pi^⁻)\simeq 1$. Thus, we would expect up to $L_{int}\times \bar{\epsilon}\times \mathrm{Br}\times(d\sigma_{\rho}/dy)\sim 3.5\times 10^4$ decays in the channel $\rho \rightarrow \pi\pi$ at CMS  and $3.5\times 10^3$ decays at ALICE per unity of rapidity. 

Let us discuss the comparison to other model predictions for $\rho$ and $\phi$ photoproduction. In Ref. \cite{Guzey:2018bay} a model based on the Glauber-Gribov approach and  hadronic fluctuations for the photon-nucleon cross section, which is in agreement with the experimental results for $\rho$ photoproduction at RHIC (AuAu) and LHC (PbPb runs) energies, has been considered. The authors provide predictions for low mass meson photoproduction in Xe+Xe UPCs. For the incoherent case, they also analyse the size of target nucleon dissociation contribution in detail. That study predicts $d\sigma_{\rho}/dy(y=0)\sim 175 \,(25)$ mb for coherent (incoherent) differential cross sections, which are somewhat higher than ours. One possible source of difference is the inclusion of low-energy photoproduction related to the secondary Reggeon exchange in the $\rho-N$ interaction. Moreover, they found $d\sigma_{\phi}/dy(y=0)\sim 17.5 \,(2.5)$ mb for coherent (incoherent) $\phi$ production, showing the same trend as in the $\rho$ case. On the other hand, the dedicated 
STARlight Monte-Carlo generator \cite{STARlight,KN} is  based on the parametrization of the forward $\gamma p \rightarrow V p$ cross section, vector meson dominance (VMD) and using DESY-HERA data to fix the $\gamma p\rightarrow Vp$ cross section. The cross-sections for coherent production on nuclear
targets are determined using a classical Glauber calculation. In general, STARlight is consistent with experimental results for coherent and incoherent $\rho$ production in UPCs in PbPb mode \cite{STARlight,RhoALICE}.  

Also a recent study considering Xe+Xe collisions has been done in Ref. \cite{Cepila:2018zky}, where photoproduction of vector mesons off proton and off nuclear targets using colour dipole model in an approach that includes
hot spots \cite{Cepila:2017nef}. The hot spots have position in the impact-parameter plane changing event-by-event and their  number depends on the energy of $\gamma p$ system. The authors have found for the cross section at midrapidity for $\rho$ production the following values: $167.5 \pm 7.5$ mb (coherent Xe+Xe) and $8.15 \pm 1.65$ mb (incoherent Xe+Xe). The lower bounds correspond to the case considering nucleons with hot spots. The results are in agreement with our predictions, mostly for the incoherent cross section. 

\begin{figure*}[t]
		\begin{tabular}{ccc}	
			\includegraphics[scale=1.0]{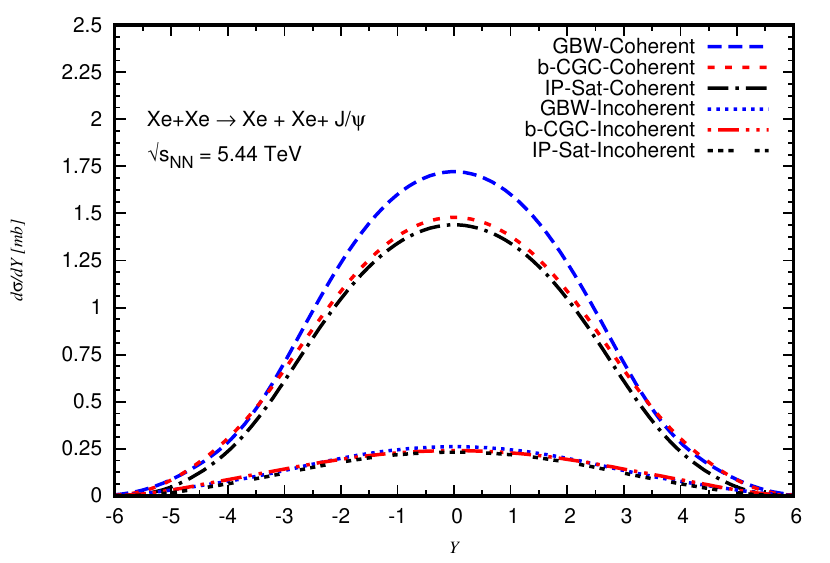} &
			\includegraphics[scale=1.0]{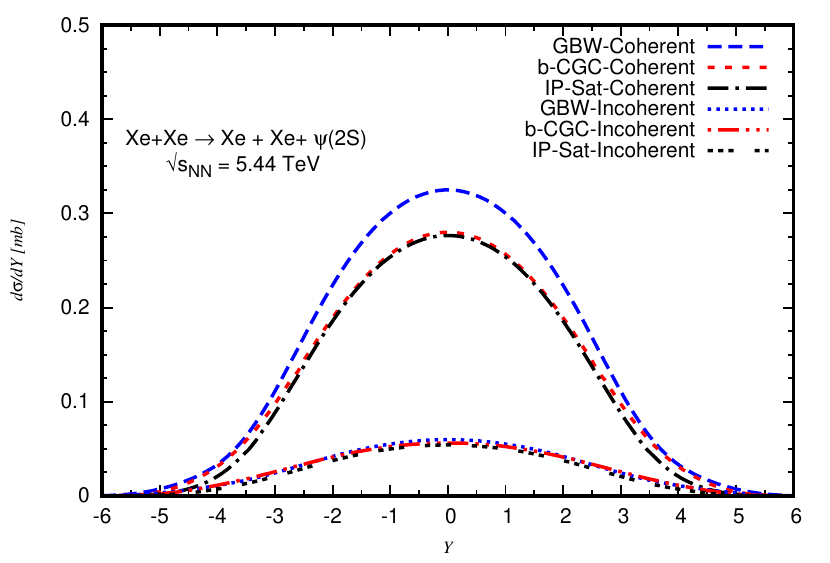} &
		\end{tabular}
			\caption{Rapidity distributions for the  coherent (upper curves) and incoherent (lower curves) photonuclear production of $J\psi$ and $\psi$(2S)  for GBW, b-CGC and IP-SAT dipole models.}		\label{fig:2.2}
	\end{figure*}
	
		\begin{figure*}[t]
		\begin{tabular}{ccc}
			\includegraphics[scale=1.0]{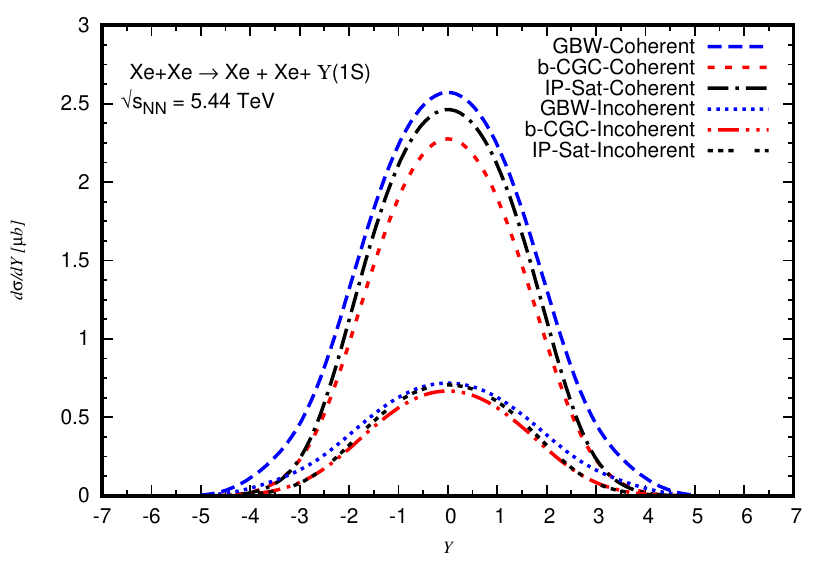}&
			\includegraphics[scale=1.0]{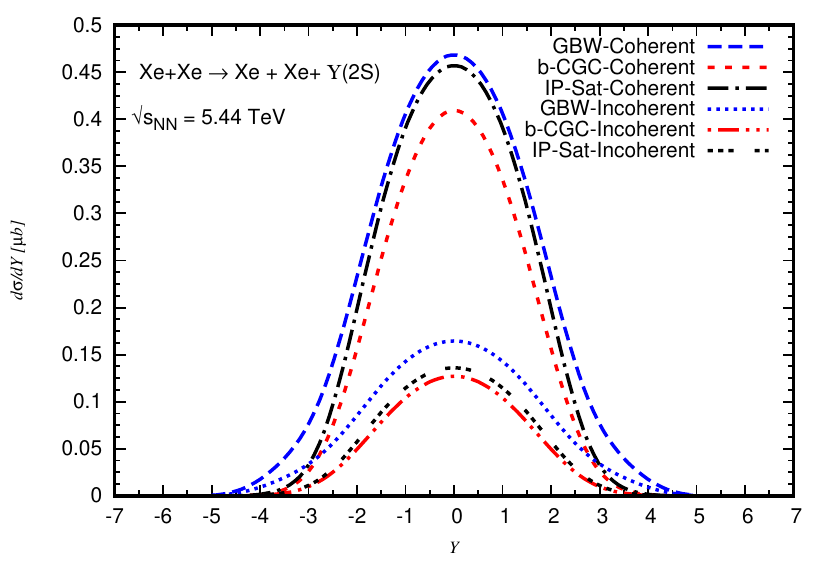} &
	    \end{tabular}
		\caption{Rapidity distributions for the  coherent (upper curves) and incoherent (lower curves) photonuclear production of $\Upsilon$(1S) and $\Upsilon$(2S)  for GBW, b-CGC and IP-SAT dipole models.}		\label{fig:2.3}	
		\end{figure*}

We now focus on the photoproduction of heavy mesons and their excited states, which it is presented in Figures \ref{fig:2.2} and \ref{fig:2.3}.  The rapidity distributions for $J/\psi$ (left panel) and $\psi(2S)$ (right panel) are shown in Fig. \ref{fig:2.2} using the same notation as the previous figure. As already mentioned, the incoherent cross sections are represented by the lower curves.  One obtains $d\sigma_{coh} /dy\, (y=0)= 1.58\pm 0.14$ mb  ($0.30\pm 0.02 $ mb) for $J/\psi\,(\psi (2S))$, whith the ratio being $\sigma(\psi(2S))/\sigma (J/\psi )$ of around 0.2. The theoretical uncertainty seems to be smaller than for the light mesons. The incoherent cross section is typically 20\% of the coherent one. Notice that the ratio $\sigma_{inc}/\sigma_{coh}$ provides further constraints on the treatment of the nuclear
modifications implemented in the different models. Now the current statistics is quite low, given an efficiency for coherent $J/\psi$ in CMS ($\bar{\epsilon}simeq 6$ \%) and ALICE  \cite{CMSJPSI} to be $\bar{\epsilon}=(\mathrm{Acc}\times \varepsilon)_{J/\psi} \sim 16$ \% and the branching ratio $\mathrm{Br}(J/\psi \rightarrow \ell^+\ell^-)\simeq 0.06$. One gets $L_{int}\times \bar{\epsilon}\times \mathrm{Br}\times(d\sigma_{\psi}/dy)\sim 42$ decays in mode $J/\psi \rightarrow \mu^+\mu^-$  at CMS  and $5$ decays at ALICE per unity of rapidity. Finally, the rapidity distributions for $\Upsilon (1S)$ (left panel) and $\Upsilon(2S)$ (right panel) are shown in Fig. \ref{fig:2.3} for completeness.

\section{Summary}
 We presented the predictions of rapidity distribution for the  exclusive vector meson photoproduction for the LHC run using Xe+Xe collisions at the energy 5.44 TeV. Predictions for the coherent (without nucleus breack up) and  incoherent cross sections are presented using the color dipole approach and Glauber-Gribov treatment of nuclear shadowing. The main focus is on the gluon saturation approach, where the main quantity is the nuclear saturation scale which is the typical momentum scale for the problem. It is considered  a consistent formalism where the wavefunction of bound states and their radial excitations are theoretically well constrained (Boosted Gaussian wavefunction). We show that the expected yields are enough to perform reliable cross section measurements for light mesons as $\rho^0$ and $\phi$. Namely, we predict up to $3.5\times 10^4$ decays in the mode $\rho \rightarrow \pi\pi$  at CMS  and $3.5\times 10^3$ decays at ALICE per unity of rapidity.  For heaviest mesons the current statistics is quite low, but the first Xe–Xe collisions have demonstrated the potential of lighter species as a path to higher hadronic luminosity.

\begin{acknowledgments}
	This work was  partially financed by the Brazilian funding
	agency CNPq.
\end{acknowledgments}

\end{document}